\documentclass[10pt,letterpaper,twocolumn]{article} 

\usepackage{ol2}
\usepackage[draft]{hyperref}
\usepackage{amsmath}

\begin{document}

\twocolumn[ 

\title{Integrated TiO$_2$ resonators for visible photonics}


\author{Jennifer T. Choy$^{1,*}$, Jonathan D. B. Bradley$^1$, Parag B. Deotare$^1$, Ian B. Burgess$^1$, Christopher C. Evans$^1$, Eric Mazur$^{1,2}$, and Marko Lon\v{c}ar$^1$}

\address{
$^1$School of Engineering and Applied Sciences, Harvard University, Cambridge, MA, 02138 USA\\
$^2$Department of Physics, Harvard University, Cambridge, MA, 02138 USA\\
$^*$Corresponding author: jchoy@fas.harvard.edu
}

\begin{abstract}We demonstrate waveguide-coupled titanium dioxide (TiO$_2$) racetrack resonators with loaded quality factors of $2.2\times10^4$ for the visible wavelengths.  The structures were fabricated in sputtered TiO$_2$ thin films on oxidized silicon substrates using standard top-down nanofabrication techniques, and passively probed in transmission measurements using a tunable red laser.  Devices based on this material could serve as integrated optical elements as well as passive platforms for coupling to visible quantum emitters.\end{abstract}

\ocis{130.3120, 230.5750.}

 ] 
Optical microresonators are ubiquitous components in optical telecommunication systems and provide a compact and efficient means for studying cavity quantum electrodynamics~\cite{vahala}.  The integration of microresonators into classical and quantum networks relies on materials that support high quality optical components on a chip-scale~\cite{lipson, hosseini}.  In the growing field of quantum photonics, single photon sources such as solid-state color centers~\cite{kurtsiefer} and emitters~\cite{bawendi,moerner} operate primarily in the visible wavelengths.  Therefore, the development of a chip-scale platform for the visible is a critical step towards the realization of quantum communication networks, and can be beneficial to classical applications such as light generation and on-chip sensing.  While gallium phosphide~\cite{rivoire}, silicon nitride~\cite{hosseini, barth, khan}, silicon dioxide~\cite{yiyang}, and diamond~\cite{faraon}, are all promising materials in this regard, certain challenges remain, including intrinsic luminescence~\cite{barth}, difficulty of generating thin membranes with low optical loss~\cite{wang}, poor refractive index contrast with the surrounding medium~\cite{yiyang}, and lower tolerance for fabrication imperfections due to inherently smaller characterisitic lengths.


Titanium dioxide (TiO$_2$) can be added to the family of viable integrated visible photonics platforms.  It is a wide-bandgap semiconductor (with a bandgap energy between 3 -- 3.5 eV, depending on the crystalline phase~\cite{campbell}) with a moderately high index for the visible wavelengths ($n$ $\approx$  $2.4$) and a wide transparency window from the near UV to the IR.  It is also naturally abundant and is compatible with a host of conventional growth techniques.  While the optical properties of TiO$_2$ have been exploited in three-dimensional photonic crystals~\cite{subramania}, gratings~\cite{alasaarela}, and waveguides~\cite{furuhashi, bradley}, planar resonator structures in this material had yet to be demonstrated.  Here we show scalable and integrable TiO$_2$ racetrack resonators, having loaded quality factors ($Q$) on the order of 10$^4$ with efficient coupling to a feeding waveguide.

Amorphous TiO$_2$ thin films of thickness 170 nm have been deposited on oxidized silicon substrates using RF sputtering of a Ti target in an O$_2$/Ar environment~\cite{bradley}.  Prism coupling experiments~\cite{bradley} indicated that the deposited films have a refractive index of 2.36 at 633 nm and propagation losses as low as 2 dB/cm.  These losses suggest that the material-limited $Q$ for optical cavities is $Q_{mat}$ $\approx$ $5\times10^5$.  The waveguide-coupled racetrack resonators used in this experiment were designed to minimize bending losses (by making bending radii $R = 30$ $\mu m$), so that additional losses would be mostly due to coupling and scattering from structural imperfections resulting from fabrication.  Additionally, the waveguide-resonator separation $g$ and coupling length $L$ were chosen to ensure efficient transfer of light signal between the cavity and waveguide~\cite{chin}.

\begin{figure}[htb]
\centerline{\includegraphics[width=6.6cm]{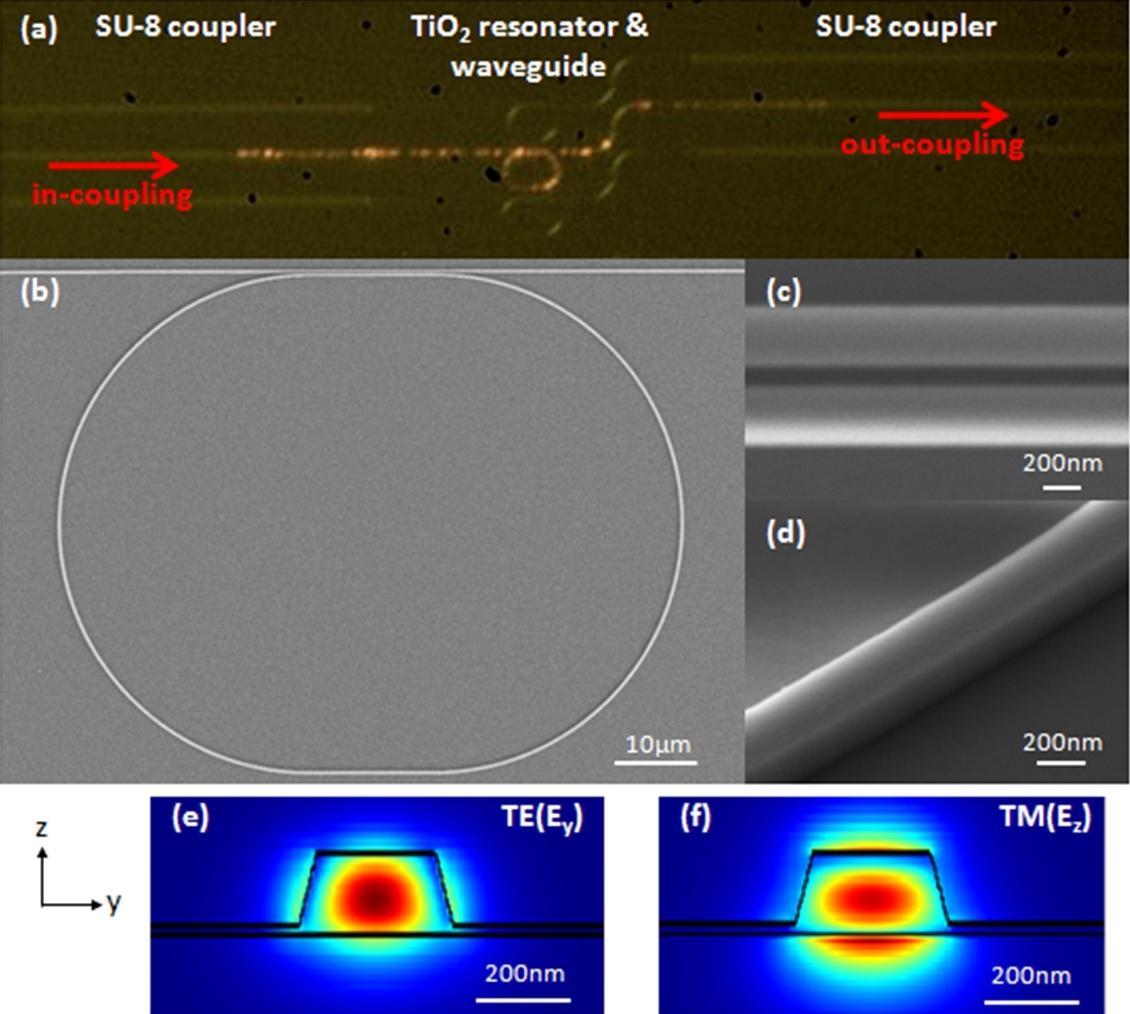}}
\caption{(a) Optical image of a set of TiO$_2$ waveguide-racetrack devices integrated with polymer pads for coupling light on- and off-chip.  Output from a HeNe laser (632nm) was fed through the middle device and coupled to both the waveguide and racetrack resonator.  SEM images of (b) a TiO$_2$ waveguide-resonator, (c) its 100-nm coupling region, and (d) etch profile taken at a 45 degree tilt.  Cross sections of the fabricated waveguide/resonator with superimposed electric field profiles of the (e) TE and (f) TM modes as calculated by FEM.  Propagation is in the $x$ direction.}
\label{image}
\end{figure}

We realized the designed structures using conventional top-down fabrication techniques which include electron beam lithography on a positive electron beam resist (ZEP 520A), electron beam
evaporation of a chromium (Cr) mask layer, metal liftoff, and reactive ion etching in a CF$_4$/H$_2$ environment at a chamber pressure of 5 mTorr.  The recipe has an etch rate of approximately 60 nm/min and leads to slightly slanted sidewalls with an angle of 75$^\circ$~\cite{bradley}.  The Cr etch mask was subsequently removed by a Cr etchant and polymer pads (SU-8 2002, with cross-sections 3 $\mu$m $\times$ 3 $\mu$m) were written using electron beam lithography~\cite{su8}.  These pads overlap the tapered region of the waveguides and extend to the edge of the chip (Fig.~\ref{image}a).  The edges of the polymer waveguides were then cleaved to facilitate in- and out-coupling of light.  The resulting resonators have smooth sidewalls, with gaps of roughly 100 nm in the coupling region (Fig.~\ref{image}b-d).  

We simulated the mode profiles of the fabricated TiO$_2$ waveguides (Fig.~\ref{image}e-f) using the finite element method (FEM).  From SEM imaging, the waveguides have a width of 250 nm on their top facets.  The etch depth is roughly 150 nm, so there is a 20-nm thick pedestal TiO$_2$ layer above the oxide film.  These dimensions support one fundamental mode in each of the transverse-electric (TE) and transverse-magnetic (TM) polarizations with respective effective indices of 1.80 and 1.61 at a wavelength of 630 nm.

\begin{figure}[htb]
\centerline{\includegraphics[width=6.9cm]{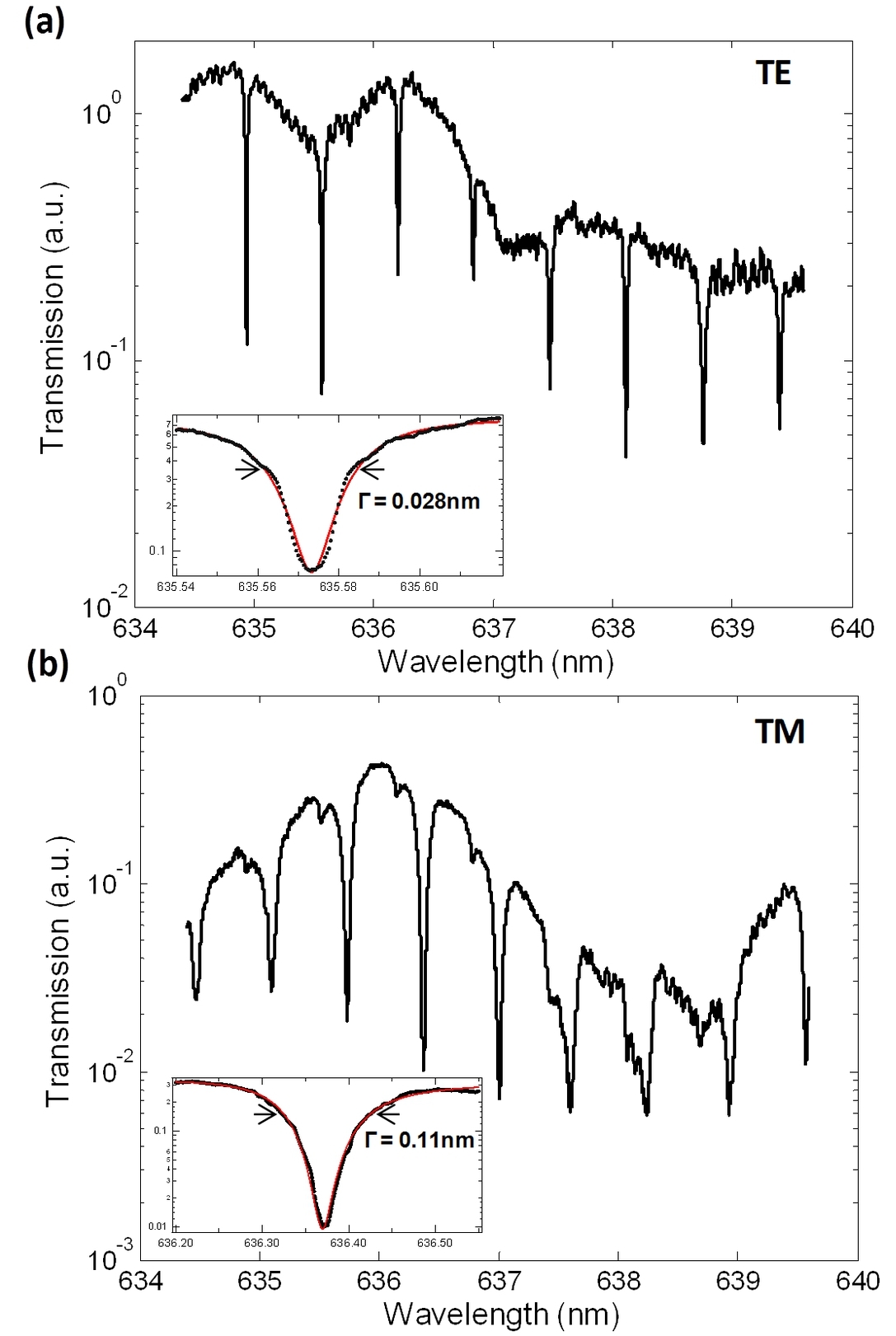}}
\caption{(a) TE-polarized transmission as a function of wavelength for a racetrack resonator with R = 30 $\mu$m and L = 15 $\mu$m. Inset: cavity mode at 634.94 nm, along with a fit to the Fano model (red).  The fitted linewidth is 0.028 nm, corresponding to a $Q$ of 22400.  (b) TM-polarized transmission as a function of wavelength for the same resonator.  Inset: cavity mode at 636.36 nm, with a fitted linewidth of 0.11nm and corresponding $Q$ of 5629.}
\label{trans}
\end{figure}

The resonators were characterized by transmission measurements using a tunable red laser with a scanning range of 634.4 -- 639.6 nm.  The output of the laser was coupled to a single mode, tapered lens fiber that was oriented such that the input signal into the sample was fixed to either the TE or TM polarization.  Light with a spot size of 0.8 $\mu m$ in diameter was focused onto the facet of the polymer pad by the tapered lens fiber and transmitted into the waveguide-resonator device.  The output was coupled through a polymer pad on the opposite end of the chip and collected by another tapered lens fiber that was connected to a high speed silicon detector.  

The resonator transmission spectra are shown in Fig.~\ref{trans}.  The transmission spectra have periodically-spaced dips corresponding to whispering-gallery modes (WGM) with either the TE (Fig.~\ref{trans}a) or TM (Fig.~\ref{trans}b) polarization.  The loaded quality factors ($Q_{loaded}=\frac{\lambda}{\mathrm{FWHM}}$, where $\lambda$ is the cavity resonance wavelength and FWHM is the full-width at half maximum at resonance) were extracted by fitting the transmission dips to the Fano model~\cite{fan}.  The fits yielded linewidths as narrow as 0.028 nm and 0.11 nm for the TE and TM polarizations (respectively) near $\lambda =$ 635 nm, corresponding to respective $Q$ values of $2.2\times10^4$ and $5.6\times10^3$.  The observed transmission drops are as large as 96\%, indicating that the resonators are nearly critically coupled. 

\begin{figure}[htb]
\centerline{\includegraphics[width=8cm]{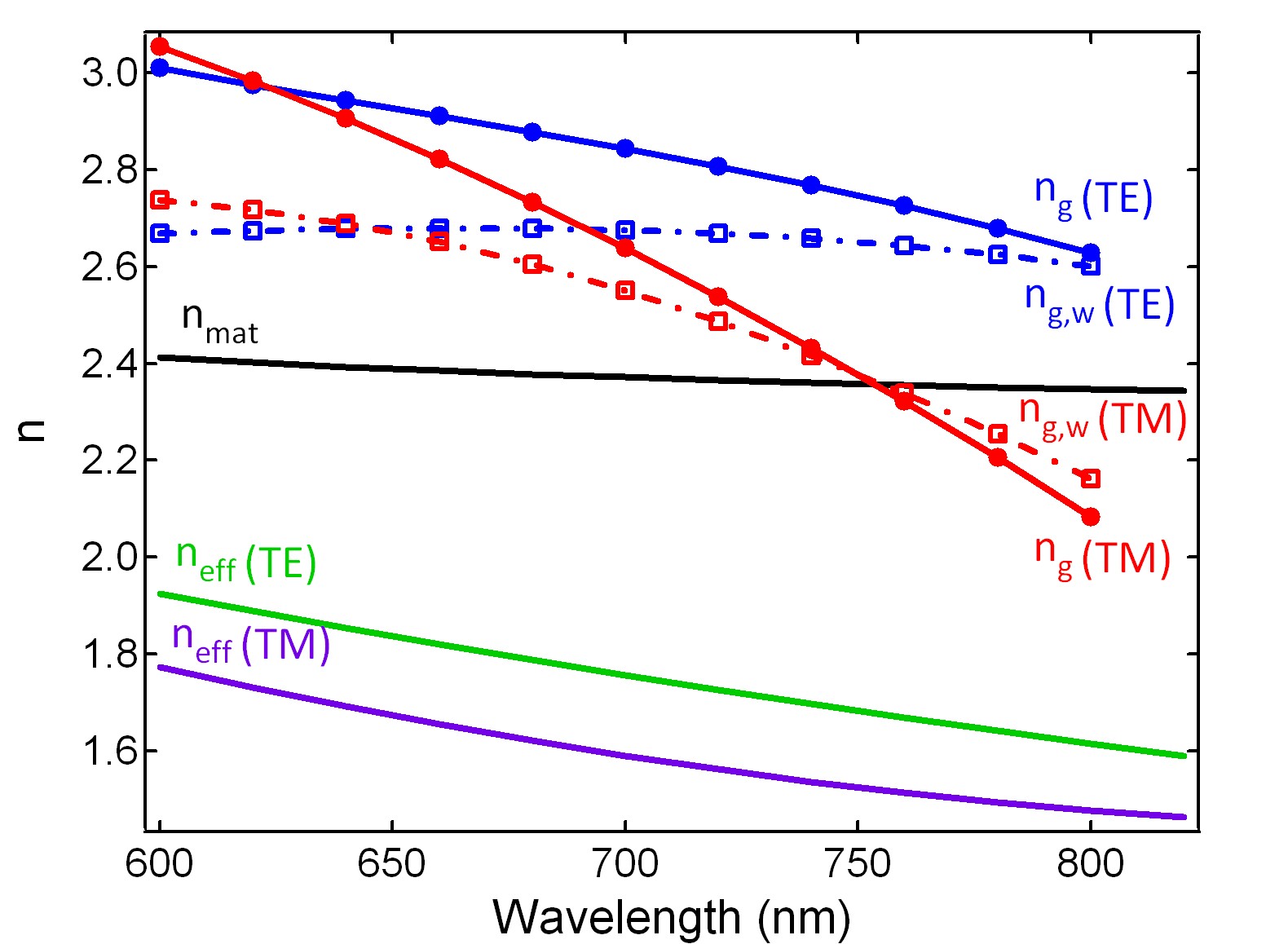}}
\caption{Calculated effective ($n_{eff}$) and group ($n_g$) indices as a function of wavelength for the TE and TM-polarized waveguide modes.  Contributions to the difference between $n_{eff}$ and $n_g$ include dispersions in the material ($n_{mat}$) and waveguide, which can be inferred by calculating $n_g$ using a fixed $n_{mat}$ of 2.4 ($n_{g,w}$).}
\label{dispersion}
\end{figure}

The transmission properties of a waveguide-coupled racetrack resonator have been well-described elsewhere~\cite{chin, zhou}.  The resonance condition is given by $n(2\pi R+2L) = m\lambda$, where $m$ is the integer mode number and $n$ is the wavelength-dependent refractive index.  Here, we have assumed that the differences between the effective indices in the coupling, waveguide, and bending regions are negligible, which was confirmed by FEM calculations.  The free spectral range (FSR) between consecutive modes $m$ and $m+1$ in such a system is described by
\begin{equation}
\mathrm{FSR}\times (n-\frac{\mathrm{d}n}{\mathrm{d}\lambda}) \approx \frac{\lambda^2}{2\pi R+2L}.
\label{fsr}
\end{equation}
The quantity $n -\lambda\frac{\mathrm{d}n}{\mathrm{d}\lambda}$ is denoted as the group index $n_g$ and takes into account dispersion in the system~\cite{guarino}.  

We used the Sellmeier coefficients obtained from spectroscopic ellipsometry measurements on our TiO$_2$ films to determine the wavelength-dependent index ($n_{mat}$).  The first order effective index ($n_{eff}$) and group index ($n_{g}$) were then computed and are shown in Fig.~\ref{dispersion}.  In the visible wavelengths, $n_g$ in TiO$_2$ waveguides can be significantly larger than $n_{eff}$ (difference of approximately 57\% for the TE mode and 70\% for the TM mode).  The greater difference observed in the TM polarization is due to larger waveguide dispersion, which can be estimated by calculation of the group index $n_{g,w}$ of the given waveguide without considering the wavelength-dependence in $n_{mat}$.  The large discrepancy between $n_g$ and $n_{eff}$ can thus be attributed to both waveguide and material dispersions, although the contribution from the latter decreases with increasing wavelengths, as indicated by the diminishing difference between the group index with and without material dispersion ($n_g$ and $n_{g,w}$, respectively).  Dispersion must therefore be considered when designing optical components in TiO$_2$ and can be exploited to generate small FSRs without significantly increasing the device footprint.  The calculated $n_g$ values are in good agreement with the experimentally obtained number, in which an FSR of $0.64$ nm (Fig.~\ref{trans}a) corresponds to a group index of 2.92 in the 635 nm to 640 nm range in the TE polarization.



Finally, the propagation loss in a critically-coupled resonator can be estimated using $\alpha_r = \frac{\pi n_g}{\lambda Q_{loaded}}$~\cite{preston}.  Based on a TE-polarized mode with a $Q$ of 22400 and transmission drop of 92\% (inset of Fig.~\ref{trans}a), the corresponding propagation loss is 28 dB/cm.  The deviation from the planar loss value can be attributed to scattering losses from surface roughness introduced by the fabrication process, which might be reduced by using a top-cladding material.

We have demonstrated planar resonators in TiO$_2$ thin films for visible light operation with efficient coupling to waveguides for delivering light on- and off-chip.  The methods and devices shown here could help advance the TiO$_2$ material platform towards integration with active emitters for novel and integrated classical and nonclassical light sources and on-chip sensing.

The research described in this paper was supported by the NSF under contract ECCS-0901469 and based upon work supported as part of the Center for Excitonics, an Energy Frontier Research Center funded by the U.S. Department of Energy, Office of Science, Office of Basic Energy Sciences under Award Number DE-SC0001088.  The devices were fabricated at CNS at Harvard University.  We thank Q. Quan, B. Hausmann, M. McCutcheon, M. Khan, F. Parsy, L. Xie, G. Akselrod, V. Bulovic, M. Pollnau, R. Jensen, L. Marshall, and M. Bawendi for useful discussions and help with the project.  J.T.C acknowledges support from NSF GRFP.

%

\end{document}